# 2.5 Dimensional Particle-in-Cell Simulations of Relativistic Plasma Collisions*

*Orestes P. Hastings IV,*

*Edison Liang*

*Rice University, Houston, TX 77005-1892*

## Introduction

We use Quicksilver, a three-dimensional Particle-in-Cell (PIC) relativistic electromagnetic code of Sandia National Laboratories, and Zohar, a 2.5 (2-D space, 3-momenta) dimensional PIC code to study relativistic collisions of electron and positron plasmas. Specifically, we use these codes to investigate two-stream instability and Weibel instability when two streams of plasmas collide head-on.

## Computational Simulation

Due to the complexity of plasma dynamics and kinetics, it is often useful in research to model plasmas through computer simulations. There are several techniques for this, but one popular one is Particle-in-cell (PIC) simulations. PIC codes have been used in the plasma physics community for several decades. The details of PIC simulations are described thoroughly in a number of textbooks and papers. (Birdsall and Langdon 1991). The general idea is that individual particles interact with only averaged electromagnetic fields createdself-consistently by currents and charge densities as well as those externally imposed. Close encounters due to coulomb collisions are ignored. Such plasmas are called "collisionless". The collisionless approximation is valid to first order only if the currents and densities are averaged over many particles in each Debye sphere (Landau and Lifshitz 1980).

## QUICKSILVER and ZOHAR

We are using QUICKSILVER, a massively-parallel, finite-difference, three-dimensional, fully-relativistic, electromagnetic, PIC code developed at Sandia National Laboratories. The information below comes from its user's manual (Coates et al 2002). QUICKSILVER can be used to simulate ion and electron diodes, magnetically insulated transmission lines, microwave devices, electron beam propagation, and high-current plasma devices. It performs time-dependent, electromagnetic, charged-particle simulations. It is written in Fortran 77, except for some low-level machine-dependent routines. It can run on both shared-memory and distributed-memory supercomputers, although it was initially designed for use on systems with a large shared central memory and multiple vector processors. QUICKSILVER uses a nonuniform finite-difference mesh with staggered full and half grids. Different regions can have varying size grids and Cartesian, cylindrical, and spherical coordinates can all be used. The QUICKSILVER field solver includes both explicit and implicit finite-difference, leap-frog algorithms. The explicit algorithm is faster but usually noisier in simulations with particles.

QUICKSILVER has a preprocessor called MERCURY that is used to set up a run from an input file. A QUICKSILVER simulation requires three basic steps: generating a QUICKSILVER input deck with MERCURY, executing QUICKSILVER, and postprocessing/visualizing the simulation. QUICKSILVER produces a number of output files including field snapshot data, particle snapshot data, and time history data.

All output of QUICKSILVER is in the form of a compact machine-portable format called Portable File Format (PFF). QUICKSILVER itself contains no graphics calls so all graphical output must be produced through the postprocessing. Initially we tried to find a way to do this with Matlab, but we learned that the best way was to license use of the PFIDL postprocessor which can be used to plot and manipulate time histories of various simulation quantities and to examine three-dimensional spatial field and particle data. It is based on the commercial IDL, which we were able to use because it is site licensed to Rice University.

For our simulations we used only Cartesian grids and the implicit field solver. All QUICKSILVER simulations were done on ADA, Rice's Cray-XDI supercomputer. It is a 632 AMD64 CPU core machine with dual core 2.2 GHz AMD Opteron 275 CPUs and with 1 MB L2 cache. Each core has 2 GB of memory, and each node has two or four cores on it, with a total of 8 GB of RAM. ADA is running SuSE 9.0 Linux and the 2.6.5 kernel.

ZOHAR is a 2.5-D PIC code developed by Langdon and Lasinski (1976) at Lawrence Livermore National Laboratory. The physics and algorithm are similar to those described in Birdsall and Langdon (1991). We use the serial version of ZOHAR mainly to calibrate results against QUICKSILVER. We ran ZOHAR on the LLNL open cluster GPS.

**Two-Stream Instability**

A popular test problem for plasma simulations is the (electrostatic) one dimensional two-stream instability. The two-stream model simulation consists of two opposing streams of charged particles (electron/positron, electron/election, etc). This system is unstable (Birdsall and Langdon, pp. 94-109). When two streams of charged particles move through each other a density perturbation on one stream is reinforced by the forces due to bunching of particles in the other stream and vice versa, each with one wavelength in one cycle of the plasma frequency, and it can be shown that the perturbation grows exponentially in time with a growth rate proportional to the electron plasma frequency. The physics of the two-stream model has already been studied in great detail, both analytically and numerically, in the literature (Lapenta et al 2007, Dieckmann et al 2007) so we will not go into further derivation here.

With QUICKSILVER, our 2-stream simulation box size is 0.4 cm x 0.4 cm x 30 cm, with the long dimension being along the *i*-axis (an *i, j, k* system). This region was broken into cells, so the region is 150 cells long and 2 cells wide in each direction. Two beams are injected each from the 2 cell by 2 cell end walls of the system. Such a small j-k grid suppresses the growth of the transverse instabilities such as the Weibel instabilities, so we can focus on the longitudinal two-stream instability. The particles are ejected cold with a velocity of 0.9c. We emitted equal numbers of electrons and positrons from both ends of the system, so the total beams from each side carried a zero net charge overall. Periodic boundary condition were used for the *j* and *k* direction and an electric conducting boundary condition was used for the *i* direction.

Initially the grid begins completely empty. The time step is $2.5 \times 10^{-12}$ seconds and at each time step 100 positrons and 100 electrons are emitted from each end wall area, so there are 800 particles emitted into the system at each time step. We allowed the system to run for 510 step ($1.275 \times 10^{-9}$ seconds) so the maximum number of particles present in the system at the end was 408,000.

In Figure 1 and Figure 2 below we plot the spatial distribution of the positrons, before and after the particles begin to collide. The electron distribution would look the same since the electrons and positrons are being emitted together.

Figure 1: The positrons just before collision

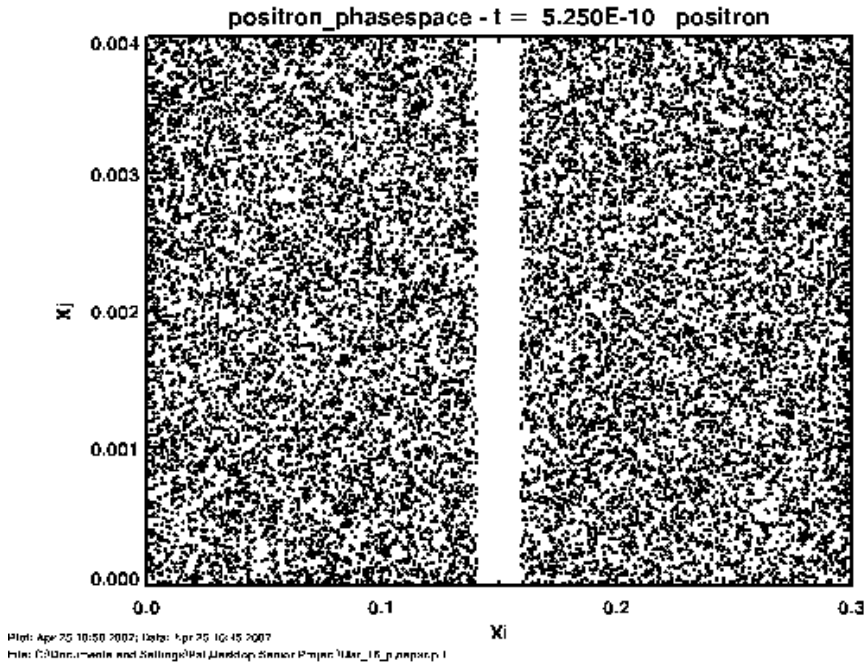

Figure 2: The positrons after collision

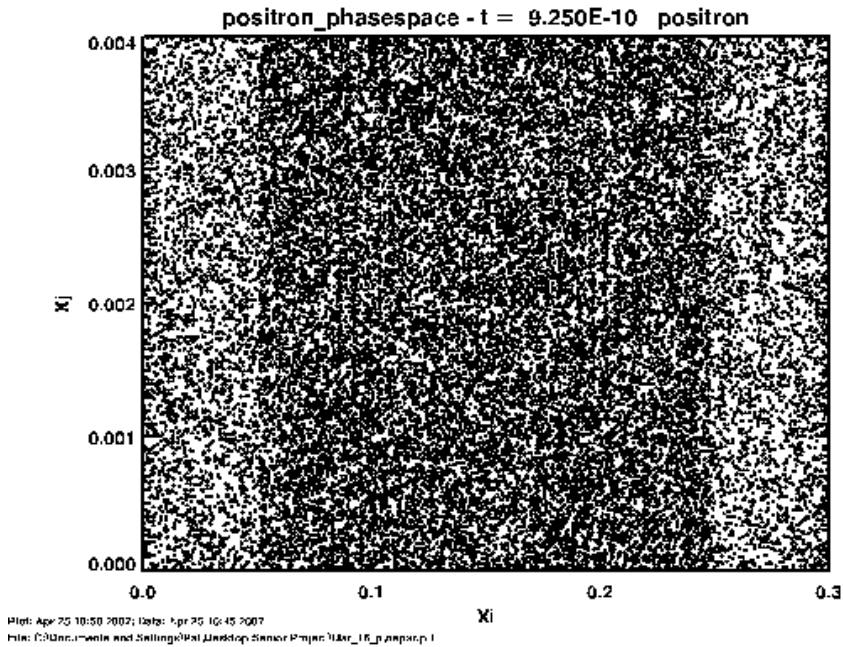

As predicted for this simulation we are able to see the ensuing two-stream instability. We can see this with phase space plots of *i*-momentum vs. *i*-position.

Figure 3: The initial momentum is 6.19 x $10^8$ (mc=$3 \times 10^8$ in QUICKSILVER units) and before the particles collide there is no real change or spread in the momentum distribution.

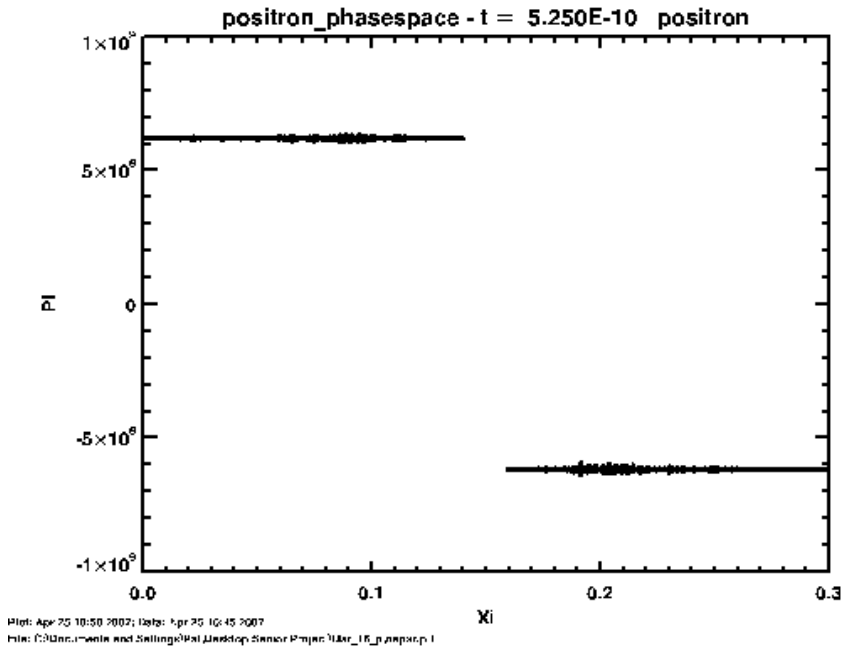

Figure 4: Right after the particles begin to collide there is some perturbing of the momentum.

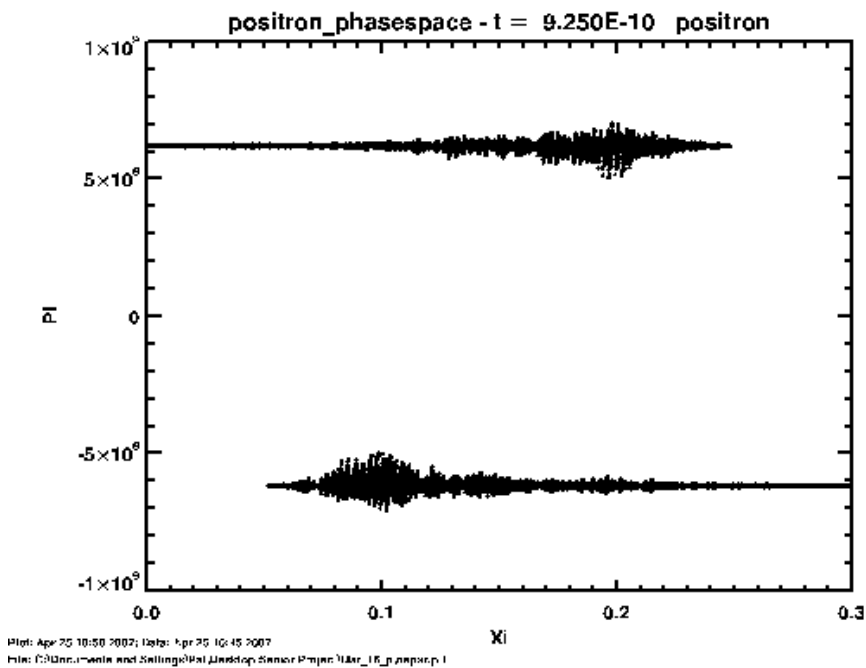

Figure 5: This effect increases exponentially.

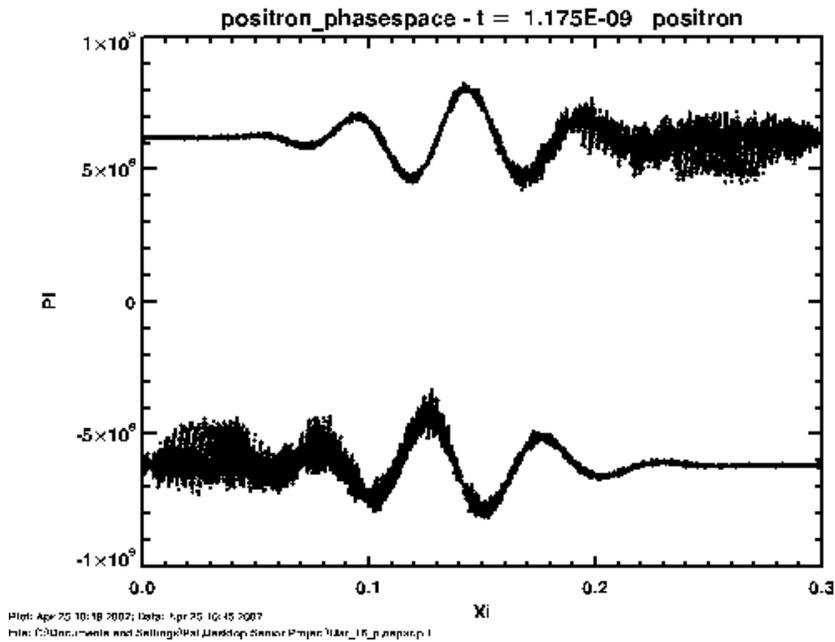

Figure 6:

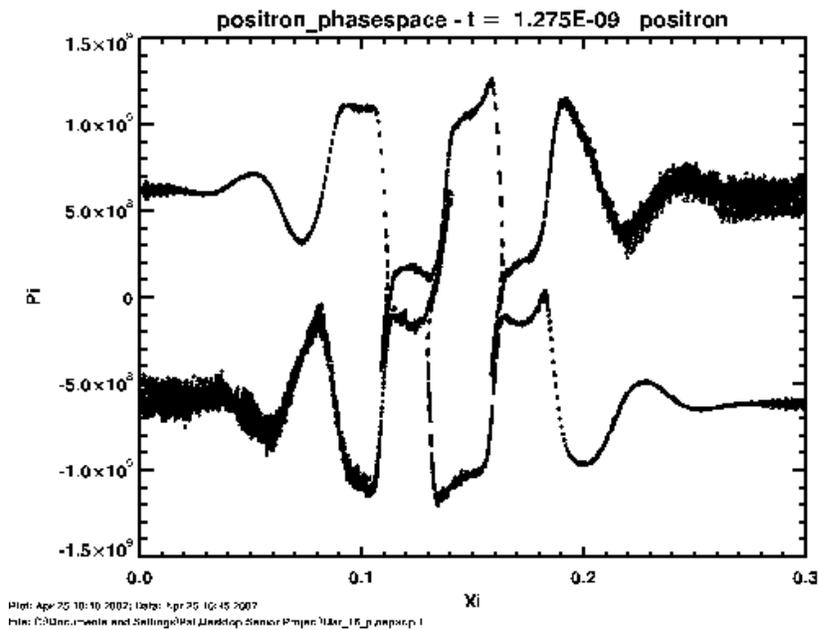

From these we see interesting sinusoidal oscillation behavior in the momentum along the i-axis due to longitudinal plasma oscillations. As expected, the initial momentum in j and k axis is small because the particles are emitted only in the i-direction. But at a later time period we found the momentum in the j and k direction to be on the same order as the momentum in the *i*-direction due to isotropization by electrostatic turbulence.

To observe the growth rate of the exponential perturbation, we can plot the magnetic and electric field energies, shown in figure 7.

Figure 7: Field Energies in the two-stream instability.

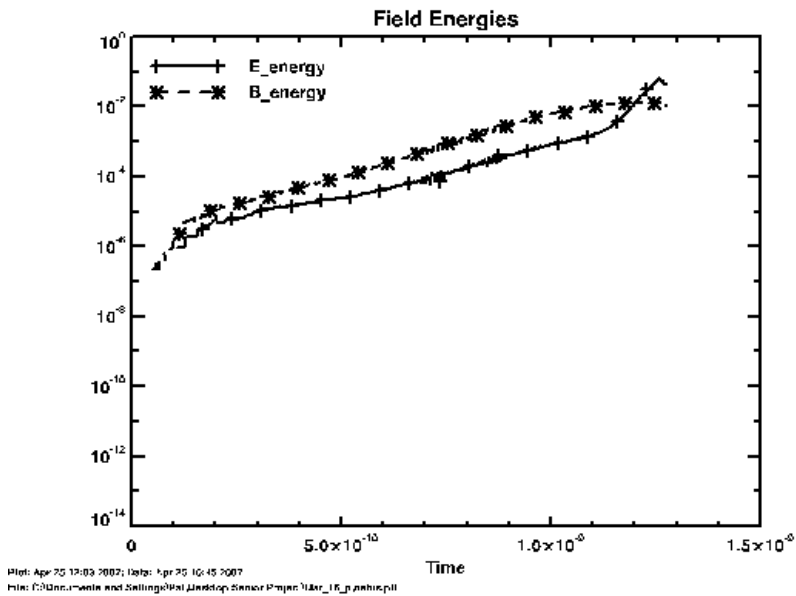

Using the ZOHAR code we also studied the growth of 2-stream instability in the ultra-relativistic regime from the collision of two electron beams of initial Lorentz factor = 10, with a neutralizing immobile ion background. We confirm the previous result of other groups (Lapenta et al 2007, Dieckmann et al 2006) that a power-law energy spectrum of nonthermal particles is produced at early times with a slope ~ -2.7 (Fig.8a). However, when we ran the code much longer, the power-law tail steepens to a slope of -3.5 and the low energy electrons are thermalized into a quasi-Maxwellian (Fig.8b).

Fig.8a Electron spectrum due to 2-stream instability at early times exhibit a power-law slope of ~ -2.7.

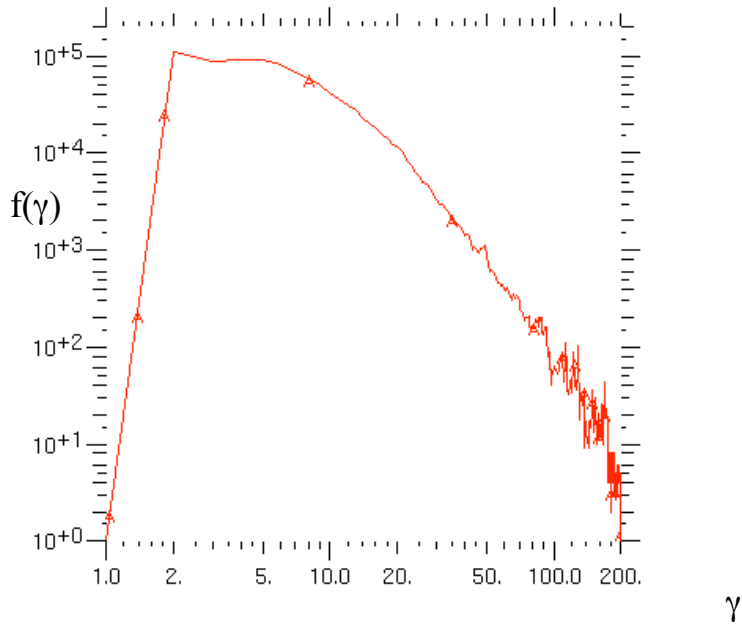

Fig.8b Electron spectra at late times exhibit a power-law tail of slope ~ -3.5 plus a quasi-Maxwellian low energy component.

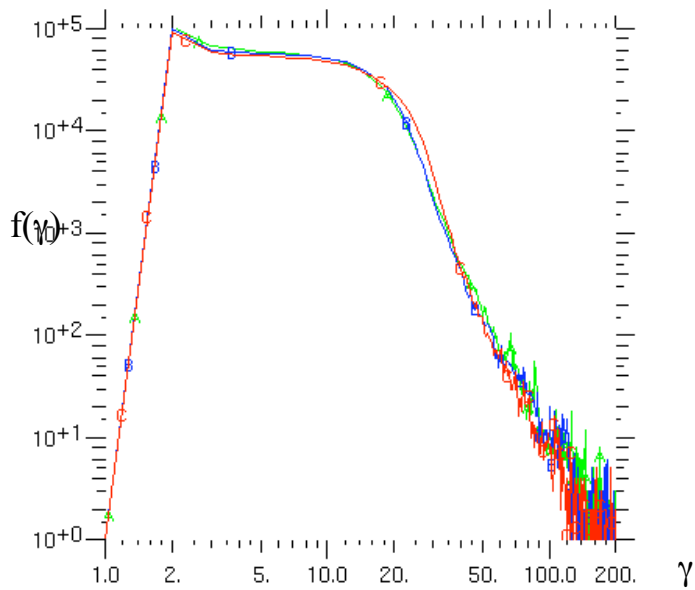

**Weibel Instability**

A second part of our project was to study Weibel instability. While two-stream instability is a one-dimensional effect, Weibel instability is strictly a multi-dimensional effect in the current configurations. Weibel instability has been studied since the 1950's (Weibel 1959, Fried 1959). Weibel instability is manifested through a bunching of currents which generate macroscopic magnetic field.

To try to observe Weibel instability in 2-D we increased the size of our simulation system along the *j*-axis to 30 cm. So our simulation system was now of size 0.2 cm x 30 cm x 30 cm. The cells maintained their same size, so now the system was 2 x 150 x 150 cells large. Again particles were still emitted from the *j-k* faces of the system. Along each cell on the face 100 positrons and 100 electrons are emitted at each timestep as before and we ran up to 900 time steps, so the maximum number of particles present in the system was $3.9 \times 10^7$.

The particle plots were similar to what we observed before. Figures 9 through 11 are plots of particle position in the *i-j* plane, and the figures 12 through 15 are the *i*-position vs. *i*-momentum.

Figure 9: *j*-position vs. *i*-position

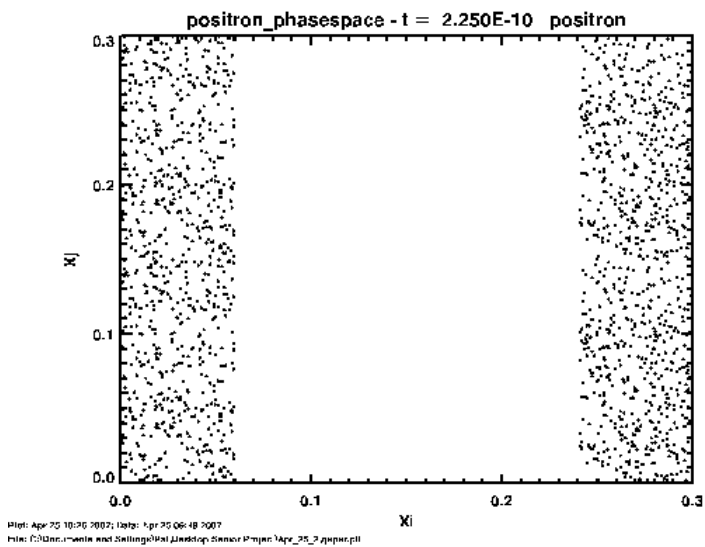

Figure 10

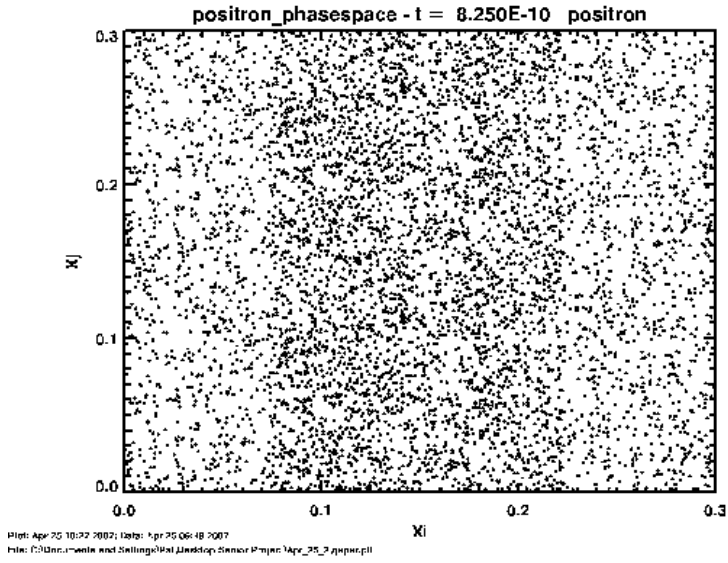

Figure 11

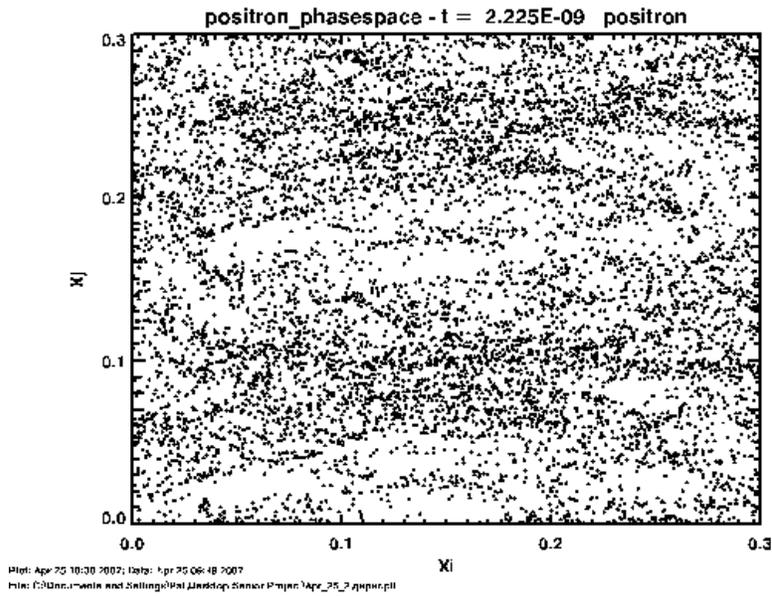

Figure 12: *i*-momentum vs *i*-position

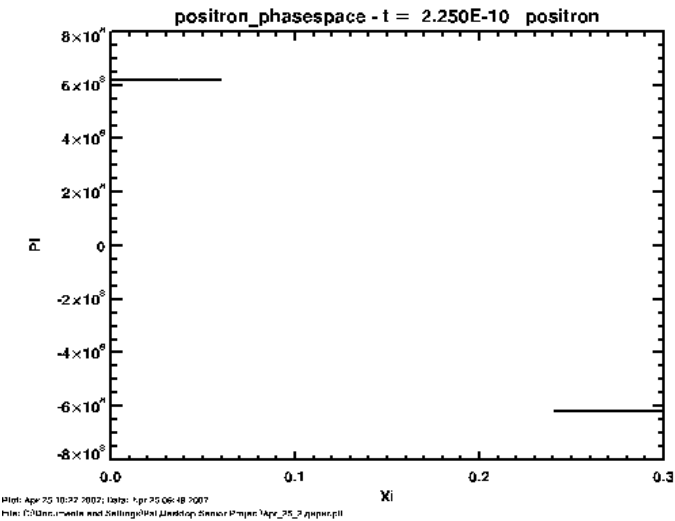

Figure 13

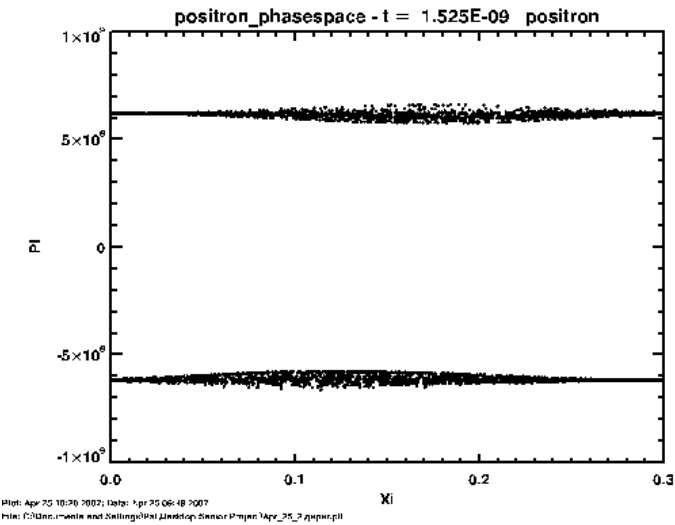

Figure 14

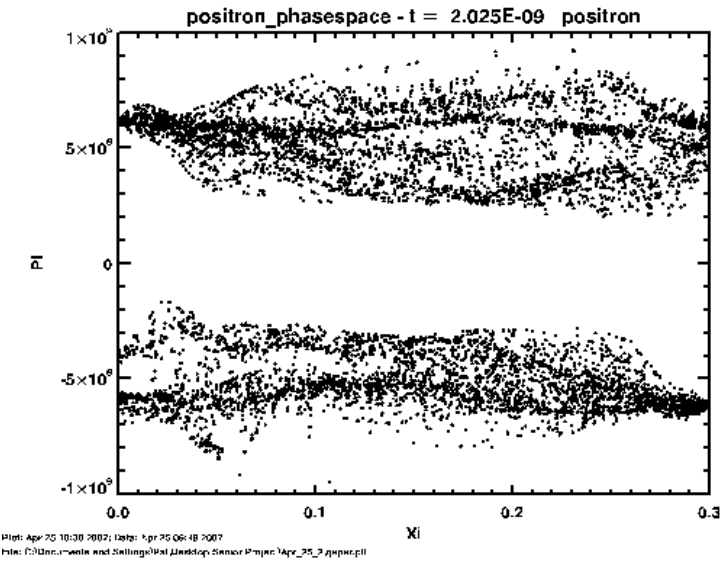

Figure 15

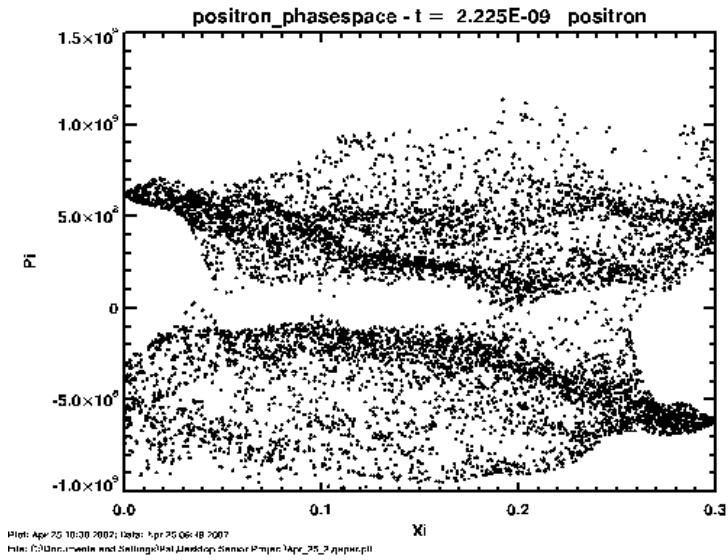

But here we see the growth of the magnetic field, something we did not see in the 1-dimensional 2-stream simulation. At first the magnetic field is extremely small. But over time this begins to grow and form fields going in and out in the k-direction, implying that sheets of current are forming in the *j-k* plane. This growth is shown in figures 16 through 20.

Figure 16

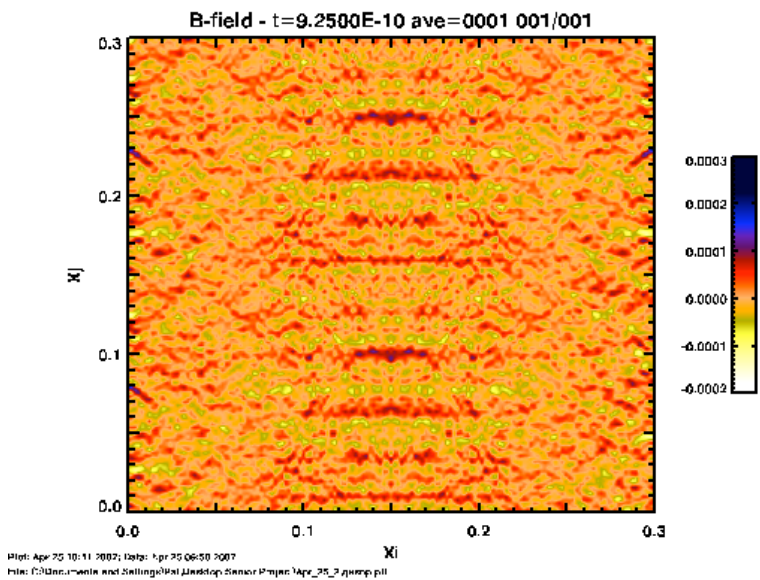

Figure 17

Figure 18

Figure 19

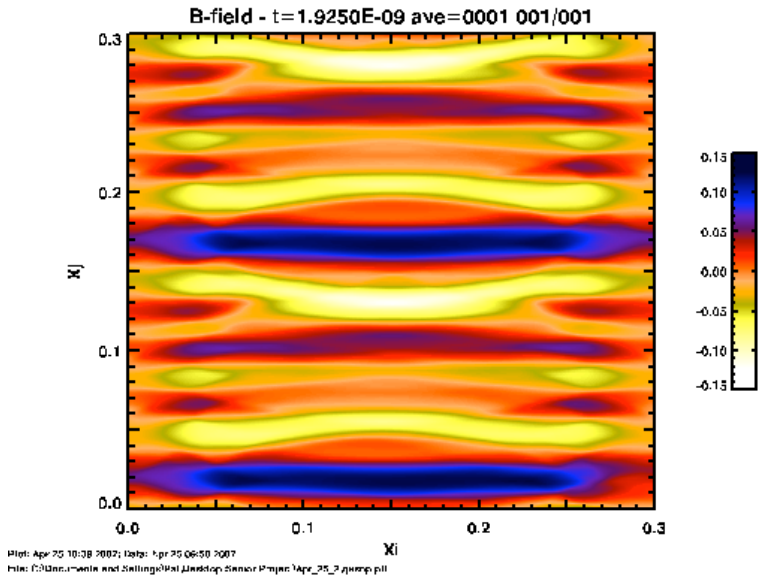

Figure 20

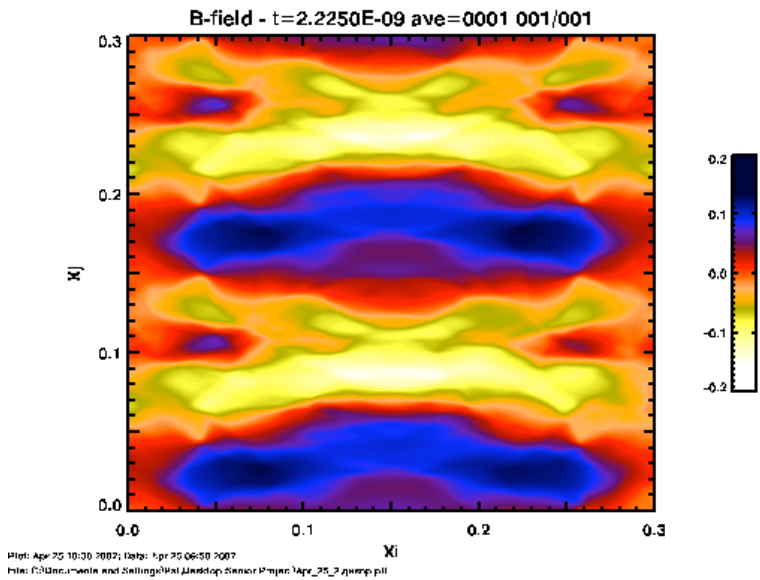

By the time we reach Figure 20 some of the current sheets have joined together, forming large islands of magnetic field. To see the exponential growth rate, we can plot the field energies. The linear growth rate is on the order of 0.1 which is similar to the analytic results of recent papers on relativistic Weibel instability. (Yoon 2007). We note that the saturation level for the magnetic field energy is greater than for the electric field energy, which is expected for Weibel instability, but not for the two-stream instability.

Figure 21: Field Energies in Weibel instability

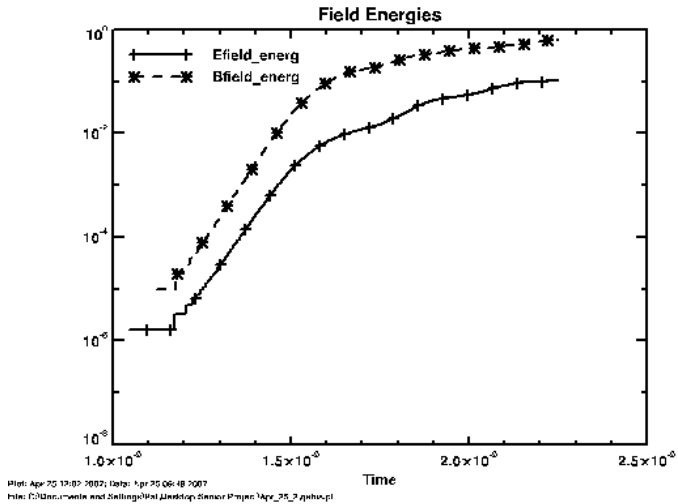

Using the 2.5-D ZOHAR code we have also studied the evolution of particle energy distribution function caused by the Weibel instability for the head-on collision of two electron-positron beams at Lorentz factors of 15. Unlike the 2-stream instability case, here we only observe a thermal Maxwellian distribution function at late times, with no trace of any power-law tail (Fig.22).

Fig.22 Time evolution of the electron spectra when two e+e- beams collide at Lorentz factors of 15. Blue spectrum shows the early spread of the electron energy from a delta function peaked at 15. Green, red and cyan spectra show the late-time asymptotic Maxwellian distribution.

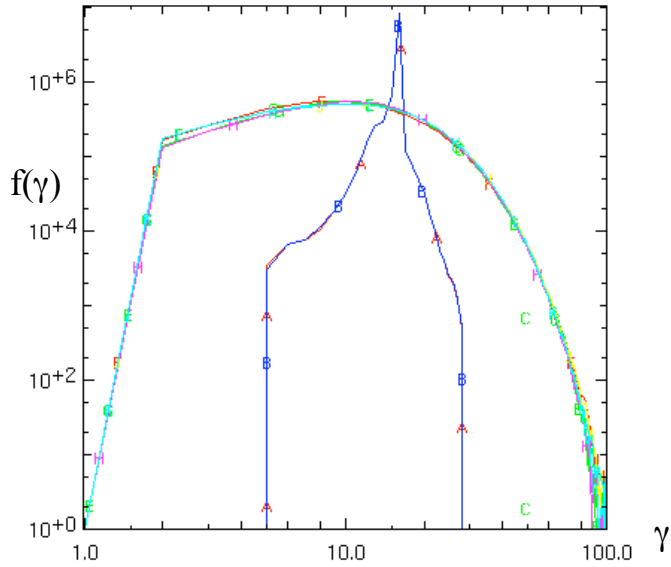

## Conclusions

We have successfully performed PIC simulations using the massively parallel QUICKSILVER code and the ZOHAR code. First by shrinking the y-z grid for streams moving in the x-direction, we were able to suppress the transverse Weibel instability and observe the longitudinal electrostatic plasma oscillations of the two-stream instability. Then by increasing the grid in the y-direction as well, we observed the Weibel instability, which was manifested in currents that generate magnetic fields observed in Figures 16-20. Our numerical instability growth rate was in basic agreement with the analytical growth rate for relativistic Weibel instability. In the electrostatic 2-stream instability case, we obtain at early times a power-law electron distribution with slope ~ -2.7, but at late times a power-law tail of slope ~ -3.5 plus a low energy Maxwellian. In the Weibel case we obtain only a Maxwellian distribution without any power-law tail at late times.

The astrophysical implications are manifold. For example, this may suggest that if relativistic collisionless shocks are primarily mediated by Weibel-induced magnetic turbulence, it may be difficult to produce nonthermal power-law electron spectra. On the other hand if Weibel instability is somehow suppressed, and the relativistic plasma collisions dissipate their kinetic energy via electrostatic turbulence, then nonthermal power-law electrons will be more easily produced. There are many known situations where Weibel is suppressed, such as the presence of longitudinal magnetic field or high electron temperature.

## Acknowledgements

We thank the staff of Sandia Laboratories: M. L. Kiefer, T. D. Pointon, D. B. Seidel, and L. P. Mix for help with this work and B. Langdon of LLNL for the use of Zohar code. This work was supported in part by the Rice Computational Research Cluster funded by NSF under Grant CNS-0421109, and a partnership between Rice University, AMD and Cray. This work was also partially supported by NSF grant AST0406882.

## References

1. C. K. Birdsall and A. B. Langdon, *Plasma Physics via Computer simulation*, McGraw-Hill Book Company, 1985.

2. F. F. Chen. *Introduction to Plasma Physics and Controlled Fusion*. Second Ed., 1984 Plenum Press, New York.

3. R.S. Coats, M. L. Kiefer, M. F. Pasik, T. D. Pointon, D. B. Seidel, QUICKSILVER Users Guide, Sandia National Laboratories, 2002

4. M.E. Dieckmann, P.K. Shukla and L.Drury, Mon. Not. R. Ast. Soc. 367, 1072, (2006).

5. B. D. Fried, Phys. Fluids, **2**, 337 (1959)

6. L. Landau and E.M. Lifshitz,, *Classical Theory of Fields* (Pergamon, London, 1980).

7. B. Langdon and B. Lasinski, *Meth. In Comp. Phys*. 16, 327, ed. Killen, J. et al (Academic, NY, 1976).
8. G. Lapenta, et al., Astrophys. J. 666, 949 (2007).
9. P. H. Yoon, Physics of Plasmas, **14**, 024504 (2007)
10. E. S. Weibel, Phys. Rev. Letters, **2**, 83 (1959)